\documentstyle[preprint,aps,prb]{revtex}
\begin{document}
\preprint{Phys.\ Rev. B {\bf 51}, 2 188 (1995)}
\draft
\title{Thermal diffusion of a two layer system}
\author{G. Gonz\'{a}lez de la Cruz and Yu.\ G. Gurevich\cite{byline}}
\address{Centro de Investigaci\'{o}n y de Estudios Avanzados  del  Instituto
Polit\'{e}cnico Nacional, Apartado Postal 14-740, 07000 M\'{e}xico D.F.,
M\'{e}xico
}
\date{\today}
\maketitle
\begin{abstract}
In this paper thermal conductivity and thermal diffusivity of  a  two  layer
system is examined from the theoretical  point  of  view.  We  use  the  one
dimensional heat diffusion equation with the  appropriate solution  in  each
layer and boundary conditions  at  the  interfaces  to  calculate  the  heat
transport in this bounded system. We also consider  the  heat  flux  at  the
surface of the samle  as  boundary  condition  instead  of  using  a  fixed
tempertaure. From this, we obtain an expression  for  the  efective  thermal
diffusivity of the composite sample in terms of the thermal  diffusivity  of
its constituent materials whithout any approximations.
\end{abstract}
\pacs{44.30.+v, 44.10.+i}

\section{Introduction}
     During  the  last  decade,  several  methods  have  been  developed  to
determine thermal diffusivities and conductivities with  high  precision  by
means of photothermal effects.\cite{Vargas88}  The most widely used method is
based on
the
photoacoustic effect. The principle of the effect is that when a sample in a
closed  cell  is  illuminated  by  light  modulated  or  chopped  at   audio
frquencies,  an  acoustic  signal  is  produced. The  application   of   the
photoacoustic effects to the measurement of thermal diffusivities  for  thin
films has been made for Adams and Kirkbright.\cite{Adams77}  They have  used
the  method called rear-surface illumination. When the  rear  surface  of  a
sample  is
illuminated with the chopped light beam, heat oscillations generated therein
propagate from the surface into the sample.  Pressure  oscillations  of  the
same frquency are induced in the gas chamber by the temperature oscillations
at the surface interface between the sample and the gas, where they  can  be
detected by a microphone. The photoacoustic signals  obtained  have  certain
phase  shift  relative  to  the  signal  detected  from  the   front-surface
excitation. Besides that, since the phase  shift  of  the  signal  does  not
depend on the optical properties of the sample, it  is  simpler  to  extract
information on the thermal diffusivity from the experimental results.  Phase
shift lag masurements are then most suitable  for  determining  the  thermal
diffusivity of the material. Charpentier et al.\cite{Charpentier82}  made  an
analysis of the pressure variations considering rear and front-surface
excitation and gave a formula relating the lag shift of  the  photoacoustical
signal  using  high modulation frequency.  The  refinement  of  the  theory,
in  terms  of  the Rosencwaig and Gersho theory\cite{Rosencwaig76}  was given
by Pessoa et al.\cite{Pessoa86} , they  showed the relative phase lag
exhibits no explicit dependence on the absorbed power and
surface conditions so that a  single  modulation frequency  measurement  is
sufficient to determine the thermal diffusivity.

     In  recent  years,  has  been  some  interest  in  study  the   thermal
characterization  of  two-layer  system  of  variable  thickness  using   the
photoacoustic effect. Tominaga and Ito\cite{Tominaga88}  used the
Rosencwaig-Gersho model
to
a two-layer system under rear illumination and looking at  the  phase  angle
behaviour as a function of the modulation frequency.  They  showed  that,  a
high modulation frequencies, the rear-illumination phase angle depends  upon
a critical frequency above which one  of  the  materials  becomes  thermally
thick.  From  the  analogy  between  thermal  and   electrical   resistances
widely used in heat-transfer problems,\cite{Karplus59} Mansanares et
al.\cite{Mansanares90}
calculated the effective thermal diffusivity of the two layer system as a
function  of  the
filling fraction of the composite  sample  and  the  ratio  of  the  thermal
conductivities of each material. Recently Christofides  and  Seas\cite{Seas}
made
an extension of the  theoretical  model  on  photopyroelectric  spectroscopy
of solids\cite{Mandelis85} to investigate the optical and thermal properties
of  a
two-layer sample. They examined the case where the optical absorption
coefficients  of the substrate and film vary in different ways, several
computer  simulations were performed in order to examine the correctness of
this model for  a  wide range of wavelengths and modulation frequencies.

     In the present paper,  we  take  a  different  approach  than  previous
investigations on the heat  transport  in  bounded  systems. We restrict  our
analysis  to  the  case  of  the  solution  of   heat   transport   equation
only considering the continuity of the heat flux at  the  interface  of  the
two-layer system. In addition, we only take into  account  the  decreasing
exponential term of the dynamical part of  the  temperature  fluctuation  in
each layer which gives the  physical  solution  of  the  heat  flux.  It  is
shown, that if the ratio of the square root of the diffusivity is  equal  to
the ratio of the thermal conductivity of each layer,  then  the  temperature
distribution is continuous at the interface, otherwise  the  temperature  is
discontinuous. In both cases, the effective  diffusivity  of  the  two-layer
system is the same.

\section{One-layer System}
     It is well known, that heat transport in solids is carried, by  various
quasiparticles  (electrons,  holes,  phonons,   magnons,   plasmons,   etc).
Frequently the interactions between these quasiparticles are such that  each
of these subsystems can have its own temperature and the physical conditions
at the boundary  of  the  sample  can  be  formulated  separately  for  each
temperature. For example, the physical conditions resulting in heat transport
by electrons and phonons are given by Granovski and
Gurevich\cite{Granovskiy75} and  those
for the transport by electrons and magnons are given in
Ref.\onlinecite{Kaganov66}, where the appropriate  boundary  conditions  are
formulated. The theory of heat
conduction in solids was developed by Gurevich and Kaganov\cite{Gurevich78}
using the two-temperature approximation. They showed that, in general, the
temperature of carriers and phonons in anisotropic semiconductors are unequal
even  in the interior of a bulk sample. The expression for the temperature
distribution for electrons and phonons was obtained in  bounded
semiconductors using the adiabatic boundary conditions. It is showed
that the accepted assumptions about the constancy of the temperature
gradient is only valid under  certain limits.14

     We  restrict  ourselves  for  definiteness  to  the   case   when   the
quasiparticle systems are electrons and phonons. Let $T_p$ the
characteristic phonon temperature. Then the momentum
$q \propto \frac{T_p}{s}$, $s$ is the sound  velocity,
sets a limit to the phase space volume occupied by phonons. Hereinafter,  we
shall consider two limiting cases:\cite{Gurevich89}

(i) Short wavelength (SW) phonons occupy a large volume in phase space. This
case is represented by the the inequality
\begin{equation}
                        2p \ll \frac{T_p}{s},
\end{equation}
where $p$  is  the  average  electron  momentum,  namely
$p \propto \sqrt {2mT_e}$ for
nondegenerate electron gas, and $p = p_f$ , the Fermi momentum, for
degenerate electron gas, here $T_e$ is the electron temperature.

 (ii) Large wavelength (LW) phonons occupy  a small volume  in  the  phase
space. This is the case when electrons interact with all phonons,
\begin{equation}
                                 2p \gg \frac{T_p}{s}.
\end{equation}

{}From Boltzman equation, it  is  known  that  the  degree  of  nonequilibrium
phonons  is  determinated  by  the  relationship   between   phonon-electron
($\nu_{pe}$) and phonon-phonon  ($\nu_{pp}$)  relaxation  frequency,
$\nu_{pe}$ determines the degree of phonon disturbance by  the  electrons
and  it  decreases  rapidly tending to zero for $q > 2p$, while the other,
$\nu_{pp}$,  describes  the  tendency of phonons to come to  equilibrium  as
a  result  of  energy  distribution.
Therefore, in the region $q > 2p$, only the following inequality holds
\begin{equation}
                             \nu_{pp} \gg \nu_{pe},\quad  q > 2p.
\end{equation}

In the region $ q < 2p $, both the inequality
\begin{equation}
                             \nu_{pp} \gg \nu_{pe},\quad  q < 2p
\end{equation}
can hold as well as the reverse inequality
\begin{equation}
                             \nu_{pp} \ll \nu_{pe},\quad  q > 2p.
\end{equation}

In the limit $ \nu_{pp} \gg \nu_{pe} $, the phonon-phonon
interations  are  more  frequent than phonon-electron collisions  and  more
efficient  in  terms  of  energy relaxation than energy transfer from the
electron to the phonon  subsystem. It should be noted  that  the  electron
and  phonon  subsystems,  generally speaking, cannot be characterized by a
single temperature. Therefore, steady state heat conduction can be described
by the following system of equations
\begin{equation}
                   {\rm div} {\bf Q}_e  = P_{ep} (T_e -T_p),\quad
                   {\rm div} {\bf Q}_p  = -P_{pe} (T_e -T_p)
\end{equation}
the term $P_{ep} (T_e -T_p)$ describes the transfer of  heat  between
electrons and phonons.  Here $P_{ep}$ is  a  parameter  proportional  to
the frequency of electron-phonon collisions ($P_{ep} = P_{pe}$) and  the
heat flux of  electron ${\bf Q}_e$ and phonon ${\bf Q}_p$  subsystems are
described by the usuall relationships
\begin{equation}
                   {\bf Q}_e  = -k_e {\rm div} T_e ,\quad
                   {\bf Q}_p  = -k_p {\rm div} T_p
\end{equation}
where $k_e$ ($k_p$) is the electron (phonon) thermal conductivity. If in
addition, we consider large specimens such that the dimension
$l \gg L^*$, where $L^*$ is  the scale of the electron-phonon energy
interaction, referred to as the  cooling length then the temperature of the
two subsystems are equal i.e. $T_e = T_p = T$.\cite{Bochkov83}
In this case we obtain, after sum Eqs.(6), the following equation
\begin{equation}
                                {\rm div} {\bf Q} = 0
\end{equation}
where ${\bf Q} = {\bf Q}_e + {\bf Q}_p$ is  the  total  heat  flux
carried by  electrons and phonons.

     Now, consider the situation represented by the  inequality  of  Eq.(5).
This is the case when phonon-phonon  collisions  alone  can  not  bring  the
phonon subsystem to an internal equilibrium.  If  Eq.(1)  is  true  i.e.  LW
phonons occupy a much smaller phase volume than  SW  phonons  then,  the  SW
system has enough time to  redistribute  the  energy  recived  from  the  LW
phonons  between  its  consituents  quasiparticles.   As   a   result,   the
distribution  function  of  SW  phonons  becomes  Planckian  (see
Ref.\onlinecite{Gurevich89}).
Then  the  electron-LW   phonon   interactions   relax   its   energy   more
efficient that the phonon subsystem and the LW phonons emitted by  electrons
of temperature $T_e$ are characterized by the same temperature
$T_e = T_p^{\rm LW}$. In this situation we also have two different
subsystems, one corresponds to the SW phonons with temperature $T_p^{\rm
SW}$ and the other one  corresponds  to  electron and LW phonon with a
characteristic temperature $T_e = T_p^{\rm LW}$ and
they  satisfay the following heat transport equations
\begin{equation}
{\rm div} ({\bf Q}_e + {\bf Q}_p^{\rm LW}) = -P_{pp} (T_e - T_p^{\rm LW}),
\qquad
{\rm div} {\bf Q}_p^{\rm SW} = P_{pp} (T_e - T_p^{\rm LW})
\end{equation}
where the term $P_{pp} (T_e -T_p^{\rm SW})$ represents the transfer of heat
from  LW phonons to SW phonons and $P_{pp}$ is calculated in
Ref.\onlinecite{Granovskiy75-2}.
If the size of the sample is greater than the cooling  length ($l \gg L_0$)
of this system then we obtain the condition $T_e = T_p^{\rm SW} = T $ and
after sum Eqs.(9), the total heat flux satisfies ${\rm div} {\bf Q} = 0$.

     When $\nu_{pe} \gg \nu_{pp}$ and Eq.(2) holds, we have  only  LW
phonons and they interact efficiently with electrons and in this case all the
quasiparticles system have the same temperature $T_e = T_p = T$ and the
equation for heat flux is similar to Eq.(8).

     From this brief discussion, we have shown that under certain conditions
on the realaxation frequency of electron and phonon subsystems  and  size  of
the sample the total system can be described by the same temperature $T$ and
for the total heat flux ${\rm div} {\bf Q} = 0$.

     So far, we have only described the  static  contribution  of  the  heat
transport i. e. the heat flux  is  independent  of  time.  However,  in  the
photoacoustic experiments, the incident radiation is modulated  on  time  by
the chopper and in this  case  it  is  necessary  to  consider  the  dynamic
contribution on the heat transport in the electron and phonon system. It  is
worth to mention that even the external perturbation depends  on  time,  the
dynamic contribution of the heat flux is only important when the  frequency
$\omega$ of the incident  electromagnetic  field  is  of  the  same  order
of the characteristic relaxation energy frequency of the electronic system
$\nu_\varepsilon$.\cite{Bass71} If $\omega \gg \nu_\varepsilon$ (high
frequency limit), in this
case the electron temperature cannot follows the variation of the field  and
assumes  an  average  value.  Since phonons can recive energy only from
electrons, the phonon temperature should also remain constant as function of
time.  Otherwise,  if  $\omega \ll \nu_{\varepsilon}$,  (low frequency
limit) the variation of the electron  and  phonon  temperature  is
quasi static. That means, the  static  quasiparticle  temperature  oscillates
with the same frquency of the radiation.

      In Eq.(8), we are not taking into account the distributed heat  source
resulting from the ligth absorption.

     Assuming that the temperature of the electron and  phonon  systems  are
equal,  which  is  usually  the  condition  in  most  of  the   photothermal
experiments the equation of heat conduction  in  solids  which  is
valid for $\omega \le \nu_\varepsilon$  can be written as
\begin{equation}
     \frac{\partial T({\bf r},t)}{\partial t}= \alpha\nabla^2 T({\bf r},t).
\end{equation}

In this equation, we are considering that the variation of  the  temperature
as function of $x$ is such that the heat conductivity $k$ is independent of
the coordinates, otherwise, we have to solve a non-liner heat equation. Here
the diffusivity $\alpha$ is given as $\alpha = \frac{k}{\rho c}$ where
$\rho$
is the density and $c$ is the specific heat of the sample. Consider the
photoacoustic  cell  geometry  for
the heat transmission  configuration  shown  schematically  in  Fig.1a.  The
temperature fluctuation is obtained from  the  solution  of  Eq.(10)  in  one
dimension. The solution $T(x,t)$ should be supplemented by boundary
conditions at $x=0$. In the photoacoustic  experiments,  the  most  common
mechanism  to produce thermal waves is  the  absorptin  by  the  sample  of
an  intensity moduladed light beam with frequency modulation $\omega \le
\nu_\varepsilon$. It is clear  that fixed the intensity of the radiation, the
light-into-heat conversion at  the surface of the sample can be written as
\begin{equation}
               \left. Q(x,t)\right|_{x=0} = Q + \Delta Q e^{i\omega t}
\end{equation}
where $Q$ is proportional to the intensity of high frequency light
($\Omega \gg \nu_\varepsilon$) and the other term represents the modulation
of this light. The  temperature is not used as boundary condition because it
is usually an unknown parameter in the experiments and besides that it is
necessary to know the  temperature on both surfaces. It is  only  important
in  thermoelectric  phenomenon  in semiconductors when the specification of
the temperatures on the surfaces of the sample must to be known.

     The general solution of the heat diffusion equation for one-layer system
is given by
\begin{equation}
             T(x,t) = T_0  + T_1 x + T_2  e^{i\omega t - \sigma x}.
\end{equation}

The parameter $\sigma$ is determined by forcing Eq.(12) to satisfies Eq.(10)
for one dimensional heat flux and is equal to $\sigma =
(1+i)\sqrt{\omega/2\alpha}$ and using the boundary condition at $x=0$, the
constants $T_1$  and $T_2$ are given by
\begin{equation}
T_1 = - \frac{Q}{k},\quad
T_2 =\frac{\Delta Q}{2k}\left[\frac{2\alpha}{\omega}\right]^{1/2}(1+i).
\end{equation}

In arriving to Eq.(12), we assume that the sample is optically opaque to the
incident light (i.e. all the incident light is  absorbed  at  the  surface),
here $T_0$ is a constant  which  cannot  be  determined  from  this
boundary conditions and it is not important in obtain the  physical  results.
It is worth to mention that the increasing exponential term with distance
in the dynamical
part of  Eq.(12)  has  not  been  considered  because  this  term  represents
a macroscopic heat flux from the lower to higher temperature region (heat
flux cannot be reflected).

     Once we know the temperature distribution in the sample, we can  assume
the acoustic piston model for evaluating the  pressure  fluctuation  in  the
cell. Acording to this model,\cite{Rosencwaig76} the ocillating component  of
the
temperature attenuates rapidly to zero with increases distance from the
sample surface.
The thin gas boundary layer at the interface is  then  to  be  acting  as  a
vibrating piston. The displacement of this piston  is  estimated  using  the
ideal gas law ($PV=Nk_B T$) for the boundary layer. As  a  result  of  this
gas piston oscillation, a pressure fluctuation $\delta P(t)$ is produced in
the cell and is given by
\begin{equation}
\delta P(t) = \frac{P_0}{T_0}\frac{\Delta Q}{2k}
\left( \frac{2\alpha}{\omega} \right)^{1/2}
\left[\cos (\omega t -\sqrt{\frac{\omega}{2\alpha}L}) +
\sin(\omega t - \sqrt{\frac{\omega}{2\alpha}L})\right]
\exp{ \left( \frac{\omega}{2\alpha} \right)^{1/2}}
\end{equation}
where $P_0$ and $T_0$ are the ambient pressure and temperature respectively.
Then Eq.(14) may be evaluated  for  the  magnitude  and  phase  of  the
acoustic pressure wave produced in the cell by the photoacoustic effect.

\section{Two-layer system}

     Let us consider the two-layer system  shown  schematically  in Fig.(1b)
consisting of a material 1 of tickness $l_1$ and of material 2 of tickness
$l_2$, both having the same cross section. Let $L = l_1 + l_2$ denote the
total sample thickness, $\alpha_i$ the thermal diffusivity and $k_i$ the
thermal conductivity of the material $i$ ($i$ = 1, 2). The system of heat
diffusion equations describing the heat  transfer  through  the  various
layers of the one dimensional photoacoustic configuration are given by
\begin{equation}
\frac{\partial T_i}{\partial t}=\alpha_i\frac{\partial^2 T_i}{\partial x^2}.
\end{equation}

The boundary conditions of the thermal diffusion equation (15) are  obtained
from the requirement of the heat flux continuity at the  interfaces  of  the
two materials and Eq.(11). The solutions of the heat transport equations can
be written as
\begin{equation}
T_1(x,t)=T_0- \frac{Q}{k_1}x + \frac{\Delta
Q}{2k_1}(1-i)\left( \frac{2\alpha_1}{\omega} \right)^{1/2}
e^{i\omega t - \sigma_1 x}, \qquad 0<x<l_1,
\end{equation}
\begin{equation}
T_2(x,t)=\theta_0+\theta_1(x - l_1)+\theta_2\exp\left[{i\omega t -
\sigma_2 (x -l_1)}\right],
\qquad l_1<x<L ,
\end{equation}
where
\begin{equation}
\theta_1= - \frac{Q}{k_2}
\end{equation}
\begin{equation}
\theta_2=\frac{\Delta Q}{2k_2} (1-i) \left(\frac{2 \alpha_2}{\omega}
\right)^{1/2} \exp\left[-\left(\frac{\omega}{2 \alpha_1}\right)^{1/2}
(1+i) l_1\right]
\end{equation}
with
\begin{equation}
\sigma_i  = \left(\frac{\omega}{2 \alpha_i}\right)^{1/2}(1 + i).
\end{equation}

Comparing temperature distribution Eq.(17) with Eq.(12) at $x=L$,  we  can
write the effective thermal diffusivity and conductivity  of  the  two-layer
system as
\begin{equation}
\frac{L}{\sqrt{\alpha}}= \frac{l_1}{\sqrt{\alpha_1}} +
\frac{l_2}{\sqrt{\alpha_2}}
\end{equation}
and
\begin{equation}
\frac{\sqrt{\alpha}}{k} = \frac{\sqrt{\alpha_2}}{k_2}
\end{equation}
or
\begin{equation}
\frac{L}{k} = \frac{\sqrt{\alpha_2}}{k_2} \left( \frac{l_1}{\sqrt{\alpha_1}}
+ \frac{l_2}{\sqrt{\alpha_2}} \right).
\end{equation}

It is worth to mention that effective thermal parameters Eqs.(21) to (23)
have been obtained whithout any approximations about the  thermal  thin  and
thickness materials,\cite{Adams77} analogy between  thermal  and  electrical
resistances used in heat transfer problems\cite{Rosencwaig76}  or some
critical frequencies
above which one of the layers becomes  thermally  thick\cite{Pessoa86} and
continuity
of  the temperature distribution at any interface.

     However, if the thermal parameters of each layer satisfy
\begin{equation}
\frac{\sqrt{\alpha_1}}{k_1} = \frac{\sqrt{\alpha_2}}{k_2}
\end{equation}
and using Eq.(21),the effective thermal conductivity can be
written as
\begin{equation}
\frac{L}{k} = \frac{l_1}{k_1} + \frac{l_2}{k_2}
\end{equation}

Then, from Eqs.(16) and (17) we obtain  that  $T_1(l_1,t)=T_2(l_1,t)$ i.e.
the temperature is continuous at the interface $x=l_1$, otherwise, in
general, the temperature will be discontinous at the interface of the two
materials  and effective thermal conductivity and diffiusivity are  given  by
Eqs.(21)  and (23). This dicontinuity in the temperature can be expressed
mathematicaly as follows:

\begin{equation}
\left. Q(x,t)\right|_{x=l_1} = \lim\limits_{\varepsilon \to 0} - k
\frac{T_2(l_1 + \varepsilon) - T_1(l_1 - \varepsilon)}{\varepsilon}=
\eta (T_2 - T_1)
\end{equation}

where $\eta$ is the surface thermal conductivity at the
interface.\cite{Bochkov83} Note
that when $\eta$ goes to infinite, since $Q$ is finite, the temperature
distribution must continuous at the interface and for the surface thermal
conductivity finite, in general, we have that $\left.T_1(x,t)\right|_{x=l_1}
\ne \left.T_2(x,t)\right|_{x=l_2}$.

\section{Conclusions}
     A theoretical analysis of the photoacustic effect on a two layer  sample
has been studied. Using the appropriate  boundary  conditions,
we obtain the effective thermal diffusivity and thermal conductivity of  the
two-layer system whithout any approximation on the thermal parameters.  The
continuity or discontinuity of the  temperature  at  the  interface  of  the
two-layer system depends on the relationship of the  thermal  parameters  of
both layers. In general, the heat flux is defined as  the  product  of  the
thermal surface conductivity and the diference of  the  temperatures  at
the interface.

     Mansanares et.al.\cite{Mansanares90}  demonstrated the uselfulness of a
single modulation frequency method for measuring the thermal diffusivity of a
solid  samples. The method consists of measure the relative phase between
the  rear-surface illumination and the front-surface illumination. Using the
thermal diffusion model of Rosencwaig and Gersho\cite{Rosencwaig76} for  the
production  of  the  photoacustic signal, the ratio of the signal amplitud
and the  phase  lag  for  rear  and front surface illumination are given as
function of the sample thickness and the sample  thermal   diffusion
coefficient.  The  theory  for  the
relative phase lag will be studied using our model in a  forthcoming  paper.
Finally, its important to mention that our  model  is  valid  for  modulated
frequency of the incident ligth $\omega$, of the same order that the
frequency of the relaxation energy between the quasiparticle systems
$\nu_\varepsilon$. In the limit,  $\omega \gg \nu_\varepsilon$ the system
can not respond to this external perturbation,  therefore,
the average in time of the dynamical part of the heat flux is neglegible and
the transferred heat is  only  static.  For  $\omega \ll \nu_{\varepsilon}$,
the heat flux is quasi-static i.e. $\partial T/\partial t=0$.\cite{Bass71}
Then
the solution of Eq.(10) for one  layer is given by $T(x,t)=C_1 +C_2 x$ and
from Eq.(11), we finally obtain the  solution  in this regime as
$T(x,t)=(C_1' +C_2'x)e^{i\omega t}$, this represents an oscillation of the
temperature distribution in every point of the layer.
\acknowledgments
This work is partially support by CONACyT.


\begin{figure}[htbp]
\caption{Geometry for (a) one-layer system and (b) two-layer system.}
\label{}
\end{figure}

\newpage
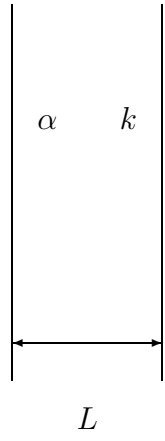
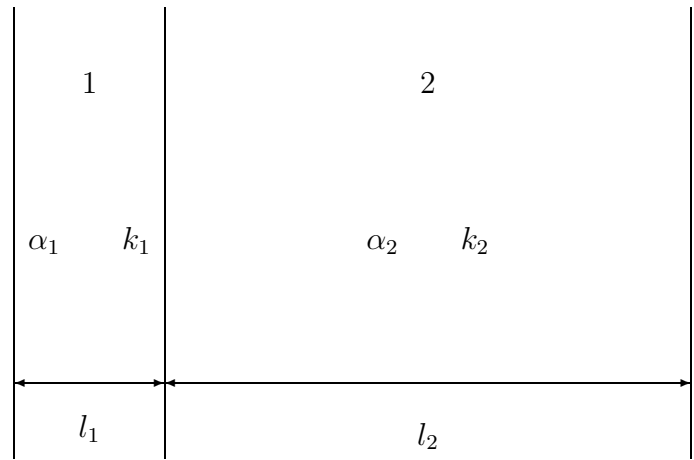
\begin{figure}
\unitlength=1mm
\special{em:linewidth 0.4pt}
\linethickness{0.4pt}
\begin{picture}(113.00,150.00)

\put(80.00,150.00){\line(0,-1){50}}
\put(60.00,150.00){\line(0,-1){50}}

\put(70.00,105.00){\vector(1,0){10.00}}
\put(70.00,105.00){\vector(-1,0){10.00}}

\put(70.00,135.00){\makebox(0,0)[cc]{$\alpha \qquad k$}}
\put(70.00, 95.00){\makebox(0,0)[cc]{$L$}}

\put( 30.00,75.00){\line(0,-1){60}}
\put( 50.00,75.00){\line(0,-1){60}}
\put(120.00,75.00){\line(0,-1){60}}
\put(85.00,25.00){\vector( 1,0){35.00}}
\put(85.00,25.00){\vector(-1,0){35.00}}
\put(40.00,25.00){\vector(-1,0){10.00}}
\put(40.00,25.00){\vector( 1,0){10.00}}

\put(40.00,65.00){\makebox(0,0)[cc]{1}}
\put(85.00,65.00){\makebox(0,0)[cc]{2}}
\put(40.00,44.00){\makebox(0,0)[cc]{$\alpha_1\qquad k_1$}}
\put(85.00,44.00){\makebox(0,0)[cc]{$\alpha_2\qquad k_2$}}
\put(40.00,19.00){\makebox(0,0)[cc]{$l_1$}}
\put(85.00,18.00){\makebox(0,0)[cc]{$l_2$}}
\put(10.00,134.00){\makebox(0,0)[cc]{(a)}}
\put(10.00,53.00){\makebox(0,0)[cc]{(b)}}
\put(20,0){\makebox(0,0){Gonzalez de la Cruz, Phys. Rev. B}}
\end{picture}
\end{figure}

\end{document}